\documentclass[pra,preprint,showkeys,preprintnumbers,amsmath,amssymb]{revtex4}

\newcommand{\funbi}[5]
{
{#1} = \left\{
\begin{array}{ll}
{#2} & \mbox{{#3}} \\
{#4} & \mbox{{#5}}
\end{array}
\right.
}

\newtheorem{theo}{Theorem}
\newtheorem{defi}{Definition}
\newtheorem{lemm}{Lemma}

\newtheorem{coro}{Corollary}

\newcommand{\proof}{{\bf Proof: }}
\newcommand{\proofend}{\hskip 15cm $\Box$}

\begin{document}


\title{Copenhagen Interpretation of Quantum Mechanics Is Incorrect
}



\author{Guang-Liang Li}
\email[]{glli@eee.hku.hk}
\thanks{Guang-Liang Li is the corresponding author.
Department of Electrical and Electronic Engineering,
The University of Hong Kong, 
Room 601, Chow Yei Ching Bldg., 
Pokfulam Road, Hong Kong, China.
Phone: (852)2857 8495, Fax: (852)2559 8738
}
\author{Victor O. K. Li}
\email[]{vli@eee.hku.hk}
\affiliation{The University of Hong Kong}


\date{\today}

\begin{abstract}
(A point-by-point response to
a comment (quant-ph/0509130) on our paper (quant-ph/0509089)
is added as Appendix C.
We find the comment incorrect.)

Einstein's criticism of the Copenhagen interpretation
of quantum mechanics is an important part of his
legacy. Although most physicists consider Einstein's
criticism technically unfounded, we show that
the Copenhagen interpretation is actually incorrect,
since Born's probability explanation of the wave function
is incorrect due to a false assumption 
on ``continuous probabilities'' in
modern probability theory.
``Continuous probability'' means a ``probability measure''
that can take every value in a subinterval of the unit
interval $(0, 1)$. We prove that
such ``continuous probabilities''
are invalid. Since Bell's inequality also assumes
``continuous probabilities'', the result
of the experimental test of Bell's inequality is
not evidence supporting the Copenhagen interpretation.
Although successful applications of quantum mechanics
and explanation of quantum phenomena do not necessarily
rely on the Copenhagen interpretation, the question
asked by Einstein 70 years ago, i.e., whether
a complete description of reality exists, still remains open.
\end{abstract}

\pacs{}
\keywords{Foundations of quantum mechanics, 
Copenhagen interpretation of quantum mechanics, Born's probability
explanation of wave function, Bell's inequality}

\maketitle


\section{\label{sec-1} Introduction}

Although Albert Einstein made many fundamental 
contributions to the development of
quantum mechanics, he remained critical to 
the Copenhagen interpretation of this
theory \cite{Pais}.
Niels Bohr was the main defender against Einstein's criticism. 
Their celebrated
debate lasted for more than a decade. 
Most physicists, however, consider this part of
the story of Einstein's life somehow ironic. 
By simply taking the quantum mechanical
description as reality itself, 
most physicists nowadays have put the issue raised by
Einstein, i.e., whether the quantum mechanical description 
of physical reality is complete \cite{Eins},
behind them. This is largely due to the result 
of the experimental test of Bell's
inequality \cite{Aspe}.
However, in contrast to commonly accepted belief, 
we shall show that the quantum
mechanical description (based on the Copenhagen interpretation) 
is actually incorrect.

Unlike Einstein's criticism, which might be due to his insistence 
on causality \cite{Pais}, the
basis of our claim above is of a technical nature. 
We prove that Born's probability explanation of 
the wave function is
incorrect (Section \ref{sec-2}), and show that
the experimental result of Bell's inequality is not
evidence supporting the Copenhagen interpretation (Section \ref{sec-3}). 

Besides the proof in Section \ref{sec-2}, Appendix \ref{app-1}
contains two more involved versions of the proof. 
Appendix \ref{app-2} discusses the hypothetical
nature of ``continuous probability'', which is an incorrect assumption
adopted in modern probability theory and causes the falsity of
Born's probability explanation.

\section{\label{sec-2} Falsity of Born's Probability Explanation}

The wave function, denoted by $\psi$, 
is the solution of Schr\"odinger's equation governing
a particle. 
According to Born's explanation, the
normalized $|\psi|^2$ is a 
``probability density function'', which implies a 
``continuous probability'', i.e., 
a ``probability measure'' whose range includes an interval.
However, assuming 
``continuous probabilities'' is a fundamental flaw in modern
probability theory.  
Actually, the range of any probability measure cannot include
intervals, so ``continuous probabilities''
are invalid. In the following, we give a
rigorous mathematical proof.

Let $(\Omega, {\cal F}, P)$ be a probability space,
where $\Omega$ is a sample space, ${\cal F}$
a collection ($\sigma$-algebra) of subsets of $\Omega$, and $P$ the
probability measure.
We shall not consider any trivial cases, such as
a probability space of
a degenerate random variable.
For a probability space of a
random variable,
we assume that 
the random variable does not take on
$\pm\infty$ as its value.

\begin{defi}
\label{de-1}
A $P$-collection is a 
nonempty family of sets in ${\cal F}$, 
such that the sets are pairwise disjoint,
and each set has a
positive
probability less than one.
\end{defi}

For a $P$-collection $F$, define

\[
\label{eq-Phi}
\Phi(F) 
= \{\gamma:\; 
\gamma = P(A), \;  A \in F\}
\]
and let $G({\cal F})$
be the
set of all $P$-collections in ${\cal F}$.
Thus, $\cup_{F \in G({\cal F})}\Phi(F)$ 
is the
union of all $\Phi(F)$, taking account of
every 
$P$-collection in ${\cal F}$.

\begin{lemm}
\label{le-1}
The set $\cup_{F \in G({\cal F})}\Phi(F)$ includes
all values 
in $(0, 1)$
that the probability measure $P$ can take.
\end{lemm}

\proof
Let $\gamma \in (0, 1)$ be a
value of $P$. There exists
$A \in {\cal F}$ with
$P(A) = \gamma$. 
Two complementary sets form
a $P$-collection
$F = \{A, A^c\}$,
with $\Phi(F) = \{\gamma, 1 - \gamma\}$.
Therefore,
$\gamma \in \cup_{F \in G({\cal F})}\Phi(F)$. 

\proofend

\begin{lemm}
\label{le-2}
A $P$-collection of a probability space $(\Omega, 
{\cal F}, P)$ is countable, i.e., there are 
countably many (finite or countably infinite)
different sets in a $P$-collection.
\end{lemm}

\proof
For a $P$-collection $F$ of the probability space, we have
\[
F = \bigcup_{n = 1}^{\infty}H_n
\]
where
\[
H_n = \{A \in F: 
1/(n + 1) < P(A) \leq 1/n\}, \;
n = 1, 2, \cdots.
\]
If $F$ is uncountable, i.e., if there are
uncountably many different sets in $F$, then
at least one of $H_1, H_2, \cdots$ must be uncountable.
Let $H_m$, where $1 \leq m < \infty$, be uncountable.
Select different sets in $H_m$, and denote the
selected sets by
$A_i, i = 1, 2, \cdots$.
Clearly, $\{A_i, i = 1, 2, \cdots\}$
is a family of pairwise disjoint sets.
Since $P(A_i) > 1/(m + 1)$ for all $A_i$, we have
$\sum_{i = 1}^{\infty}P(A_i) = \infty$.
But this is impossible. So $F$ must be countable.

\proofend

\begin{theo}
\label{th-1}
For a probability space $(\Omega, {\cal F}, P)$,
there are values almost everywhere in $(0, 1)$
that the probability measure $P$ 
cannot take. 
\end{theo}

\proof
From Lemma \ref{le-1}, we need only show
that
$\cup_{F \in G({\cal F})}\Phi(F)$ is a nullset,
i.e., a set of (Lebesgue) measure zero.
From Lemma \ref{le-2},
any $P$-collection $F$ of the probability space is
countable. So any $\Phi(F)$ is countable, and hence
is a nullset.
From the definition of a nullset,
$\Phi(F)$ can be covered
by a sequence of open intervals of arbitrarily
small total length, i.e.,
$\Phi(F)$ is a subset of the union
of the covering intervals.

On the other hand,
there is a countable base
${\cal B}$ for
the topology induced by the
usual metric on the real line (restricted on
the interval $(0, 1)$). 
Thus, for any $\Phi(F)$,
each covering interval $I$
of $\Phi(F)$ is a union of some members of ${\cal B}$.
Since the length of $I$ can be arbitrarily small,
the measure of any member 
of ${\cal B}$ contained in $I$ can also be
arbitrarily small.

Consequently, $\cup_{F \in G({\cal F})}\Phi(F)$ is a subset of a union of
the members of ${\cal B}$, such that
each member is contained in a covering interval of
some $\Phi(F)$, and
has an arbitrarily small measure.
Since any member of ${\cal B}$ is a countable
union of pairwise disjoint open intervals,
$\cup_{F \in G({\cal F})}\Phi(F)$ is covered by a sequence
of open intervals of arbitrarily small total
length.
Therefore, $\cup_{F \in G({\cal F})}\Phi(F)$ is
a nullset.

\proofend

We have also prepared two other versions of
the proof of Theorem \ref{th-1}. Both versions
are essentially the same as the proof given above, but
involve more intensive deliberation. 
Since not every reader would consider the
extra versions necessary,
we put them in
Appendix \ref{app-1}.

A ``continuous probability'' is a ``probability measure''
whose range includes an interval.
We have
an immediate consequence of Theorem \ref{th-1}.
\begin{coro}
``Continuous probabilities'' are invalid.
\end{coro}
However, modern probability theory assumes
``continuous probabilities'' in two cases.

{\em Case} I.
Consider 
a ``continuous random
variable''. 
By ``continuous''
in ``continuous random variable'', we mean ``absolutely
continuous'', i.e., the ``random variable''
has a ``probability density function'', and
the ``probability measure'' of the ``random variable''
can take
every value in a subinterval of $(0, 1)$,
as exemplified by those with
uniform, exponential, and normal distributions.
The conclusion below is an immediate corollary
of Theorem \ref{th-1}.

\begin{coro}
\label{co-1}
``Continuous random variables'' do not exist.
\end{coro}

Although ``continuous random variables'' do not exist,
there is a popular explanation of ``continuous probabilities''
based on ``continuous random variables'':
A ``continuous random variable'' is
a result of approximating 
a sum of $n$ discrete random
variables as $n$ tends to infinity.

For example,
consider $n$ independent and identically distributed 
random variables, such that the
possible values of each random variable are 0 and 1. 
As claimed by the De Moivre-Laplace limit theorem, for large $n$,
a function, defined by an integral, can approximate the
distribution of
the normalized sum of the random variables.
The sum represents the normalized number of successes in
$n$ Bernoulli trials.

Although the function given by
the integral, known as the ``standard
normal distribution'', is considered the limit
of the distribution of the normalized sum,
the approximation does not result in the so-called
``standard normal random variable''.
For any given $n$, 
the set of the possible values of 
the sum of the $n$ random variables
is $\{0, 1, 2, ..., n\}$. 
Consequently, the normalized sum for any given 
$n$ has $n + 1$ possible values. No
matter how large $n$ is, the possible values of 
the normalized sum can only form a
countable set. In other words, 
the normalized sum is a discrete random variable for
any $n$.

However, the ``normal random variable'' is a 
``continuous random variable'' with an uncountable
set of values.
The approximation is only in the sense 
that the value of the distribution of the
normalized sum and the value of 
the ``normal distribution''
can be
close. But such approximation does not
necessarily imply the closeness between the
normalized sum and the ``normal random variable''
as two functions.
The normalized sum
has a countable set as its 
range (set of possible values). 
But the 
range of
the ``normal random variable''
is an uncountable set.

{\em Case} II.
The other way leading to ``continuous probabilities''
is due to
denumerable sequences of elementary ``events''.
Such a sequence corresponds to
a decimal expansion 
of a real number in
the unit interval.
In general, the base of the expansion can be
any given integer $q > 0$. 
For simplicity and without loss of generality,
we let $q = 2$.
The following result is also immediate from
Theorem \ref{th-1}.

\begin{coro}
\label{co-2}
Let $\Omega = \{\omega_1, \omega_2, \dots\}$.
If ${\cal F}$ includes all subsets of $\Omega$, and
if $P(\omega_n) = 2^{-n}$ for all $n \geq 1$, 
then $(\Omega, {\cal F}, P)$
is not a valid probability space.
\end{coro}

``Continuous
probabilities'' are hypothetical,
introduced into probability theory through
various assumptions on sample spaces, 
$\sigma$-algebras, and ``probability measures'',
and will not exist
if the assumptions are abandoned
as they should be, since we have 
shown that such assumptions lead to
contradictions and hence are incorrect.
Actually, such assumptions are just
different ways to say that a
``probability measure'' can take
every value in a subinterval of $(0, 1)$.
Consequently, any ``proof'' of the existence of
``continuous probabilities'' (e.g., in case II)
is nothing but a tautology.
In Appendix \ref{app-2},
we show that all ``counterexamples''
to Theorem \ref{th-1} are based on
such tautology, and clarify a confusion
in some arguments of the ``counterexamples''
caused by misunderstanding on measure theory.

Although ``continuous probabilities''
are invalid, a discrete probability 
can be induced by a
``continuous probability''.
For example, consider a 
``probability density function'' $f$ with
domain $D$, which is
an interval on the
real line. Denote by $E$ a partition of
$D$, i.e., $E$ is a sequence of
pairwise disjoint subintervals $E_1, E_2, \cdots$,
such that $\cup E_i = D$. By letting
each $E_i$ in the partition
represent an elementary event
``$x \in E_i$'' with
probability $P(E_i) = \int_{E_i}f(x)dx$,
we then obtain a discrete probability.

However, once a partition $E$ is given,
for any subinterval $H$ of an
elementary event $E_i$ in $E$, we cannot
calculate the probability of $x \in H$,
since ``$x \in H$'' is neither an elementary event,
nor deducible from other
elementary events.
This leads to an uncertainty for the
induced probability itself, although
any probability is a description of some
uncertainty.
Such uncertainty in the
induced probability is inevitable. 
This is because assigning a positive
probability value to every subinterval of $D$
will lead to a ``continuous probability'', which is invalid.
Nevertheless, with finer and finer partitions,
we can decrease the uncertainty.
But since different partitions correspond to
different probability spaces,
the refinement will require infinitely
many probability distributions.

Discrete probability functions as induced above are 
actually used to calculate the
values of various probabilities and statistical quantities, 
not only in quantum
mechanics, but also in other applications 
of probability theory generally. This
explains why probability theory works well numerically, 
although ``continuous probabilities'' are invalid.

``Continuous probabilities'' and
the mathematical facts used in this paper to disprove
the existence of ``continuous probabilities''
are well-known, and can be found in standard
textbooks. For example, see 
\cite{Fell,Fell2,Foll}.
Physicists may find \cite{Roma}
more accessible. 

\section{\label{sec-3} Discussion and Concluding Remarks}

The explanation of quantum phenomena and successful 
applications of quantum mechanics
do not necessarily rely on the Copenhagen interpretation. 
For example, without Born's probability explanation, 
the solution of
Schr\"odinger's equation is sufficient 
to show the existence of discrete energy levels.

Most physicists consider the result of 
the experimental test of Bell's inequality
evidence supporting the Copenhagen interpretation. 
However, the derivation of Bell's
inequality assumes ``continuous probabilities''
\cite{Aspe}. Since ``continuous probabilities'' are
invalid, Bell's inequality itself is incorrect, 
and hence is not a valid basis for a
test. So such result is not supporting evidence 
for the Copenhagen interpretation.
On the other hand, due to the flaw in probability theory, 
the Copenhagen interpretation is incorrect, 
and hence is not eligible for a meaningful test.

After pointing out the flaw of the Copenhagen interpretation, 
we find ourselves in a
situation described by Professor Sir Hermann Bondi \cite{Bond}: 
Our work might be brushed aside
with comments like:  
``Quantum mechanics works. So there must be some fault in your
argument. Why waste time to sort it out when there 
are so many fascinating things to
be done?'' However, Einstein would definitely disagree 
with such comments. Pursuing
the truth is not a waste of time in any sense. 
The Copenhagen interpretation actually
closed the door of exploring the reality behind quantum mechanics, 
though Einstein
had tried to keep the door open. 
With this paper, we want to reopen the door.  We
conclude by citing Einstein, Podolsky, and Rosen \cite{Eins}:

``While we have thus shown that the wavefunction 
does not provide a complete
description of reality, we have left open 
the question of whether or not such a
description exists. We believe, however, 
that such a theory is possible.''

\appendix
\section{\label{app-1}More Involved Proofs}
The following two proofs
(Versions A and B) of
Theorem \ref{th-1} are essentially
the same as that we have originally given
in Section \ref{sec-2},
but involve more intensive deliberation. 
For a set $S$ on the
real line, $\mu(S)$ is the (Lebesgue) measure
of $S$. If $S$ is an interval, then
$\mu(S) = |S|$ is the length of $S$.
To avoid misunderstanding or confusion, we
first recall the definition of a set (on the
real line) of (Lebesgue) measure zero.

\begin{defi}
\label{de-2}
A set $Z$ is a set of measure zero, if for each
$\epsilon > 0$, there is a sequence of (open)
intervals $\{I_m\}$, such that 
$\cup I_m \supset Z$, and 
$\sum |I_m| < \epsilon$.
\end{defi}

For convenience of exposition, we refer to
the ``sequence'' in Definition \ref{de-2}
(i.e., in ``for each $\epsilon > 0$, there is a
sequence ...'') as a ``sequence of  intervals of arbitrarily
small total length.''
From Definition \ref{de-2}, the following is
immediately evident.

Any fixed sequence of intervals is not a sequence of
intervals of arbitrarily small total length, since
the total length of a fixed sequence of intervals
cannot be less than each $\epsilon > 0$.
Moreover, 
any fixed sequence of intervals
covering a set $Z$ is not relevant to
whether $Z$ is of measure zero.
For example, let $Z$ be a set on the real line, and
$\{I_m\}$ a fixed sequence of
intervals of total length $l$ (i.e.,
$\sum |I_m| = l$), such
that $\cup I_m \supset Z$. Although $\sum |I_m| < \epsilon$ does
not hold for each $\epsilon > 0$, and although there
are surely $0 < \delta < l$ and $I_m \in \{I_m\}$
with $|I_m| \geq \delta$, $Z$ can still be a set of
measure zero.
Actually, we have the following alternative
definition.

\begin{defi}
\label{de-3}
A set $Z$ is a set of measure zero, if for each
$0< \epsilon < l$, where $l$ is arbitrarily given, 
there is a sequence of (open)
intervals $\{I_m\}$, such that 
$\cup I_m \supset Z$, and 
$\sum |I_m| < \epsilon$.
\end{defi}

Clearly, Definitions \ref{de-2} and \ref{de-3}
are equivalent.
Let $l$ in Definition \ref{de-3}
be the total length of a fixed sequence
of intervals covering a set of measure zero. 
Since Definition \ref{de-3} (and hence Definition
\ref{de-2}) does not involve the fixed sequence of
intervals of total length $l$, and since
$l$ is arbitrary, a set of measure zero is irrelevant
to any fixed sequence of intervals.

\subsection*{Version A}
\hskip\parindent
Let $Z$ be a set on the real line with $\mu(Z) = 0$,
and ${\cal I}(Z)$ the family of sequences of 
intervals covering $Z$, i.e.,
\[
{\cal I}(Z) = \{\{I_m\}: \cup I_m \supset Z,\;
I_1, I_2, \dots \; 
\mbox{are intervals}\}.
\]

\begin{lemm}
\label{le-A}
For any decreasing sequence of positive
real numbers $\{\epsilon_m\}$
(i.e., $\epsilon_m > \epsilon_{m + 1}$ for all $m \geq 1$),
there is a sequence $\{I_m\} \in {\cal I}(Z)$, such that
$|I_m| < \epsilon_m$ for any $I_m \in \{I_m\}$.
Clearly, for any subinterval $J$ of $I_m$,
$|J| \leq |I_m| < \epsilon_m$.
\end{lemm}

\proof
Assume that the lemma is false. There is then a decreasing
sequence of positive real numbers $\{\epsilon_m\}$,
such that any $\{I_m\} \in {\cal I}(Z)$ contains
some $I_m$ with $|I_m| \geq \epsilon_m$, where
$\epsilon_m \in \{\epsilon_m\}$. Write
\[
M = \sup\{m: |I_m| \geq \epsilon_m,
|I_j| < \epsilon_j, j = 1, 2, \dots, m - 1,
\{I_m\} \in {\cal I}(Z)\}.
\]
Since $M = \infty$ implies the existence of
$\{I_m\} \in {\cal I}(Z)$ 
with $|I_m| < \epsilon_m$ for all $m \geq 1$,
we have $M < \infty$.
Thus, for any $\{I_m\} \in {\cal I}(Z)$,
$\sum |I_m| > \epsilon_M > 0$.
As a result,
\[
\mu(Z) = \inf\left\{\sum |I_m|: \{I_m\} \in {\cal I}(Z)\right\}
\geq \epsilon_M > 0.
\]
We see a contradiction. Therefore, the lemma is true.

\proofend

From Lemma \ref{le-A}, given $P$-collection $F$, for
any decreasing sequence of positive real numbers
$\{\epsilon_m\}$, we have $\{I_m\} \in {\cal I}(\Phi(F))$
with $|I_m| < \epsilon_m$ for any $I_m \in \{I_m\}$.
Evidently, $\epsilon_m$ can be arbitrarily small, i.e.,
$\epsilon_m$ can be less than any $\epsilon > 0$, for
all $m \geq 1$ (e.g., we may let 
$\epsilon_1 < \epsilon$).
On the other hand, 
Lemma \ref{le-A} applies in particular 
if we require
every $\{I_m\} \in {\cal I}(\Phi(F))$ to be
a sequence of open intervals.
In this case,
for any $\{I_m\} \in {\cal I}(\Phi(F))$,
$\cup I_m$ equals a countable union of members of the
countable base ${\cal B}$.
Let $B_i(F)$ be the $i$th member in the union of
the members of ${\cal B}$ for the $P$-collection 
$F$, 
and ${\bf N}$
the set (or a finite subset) of positive integers. 
Thus, for each $P$-collection $F$, we have
$\Phi(F) \subset \cup_{i \in {\bf N}}B_i(F)$, and
\begin{equation}
\label{eq-A}
\bigcup_{F \in G({\cal F})} \Phi(F) 
\subset \bigcup_{F \in G({\cal F})}\bigcup_{i \in {\bf N}}B_i(F).
\end{equation}

For simplicity and without loss of generality, let
each member of ${\cal B}$ be an open interval.
For example, ${\cal B}$ can be the family of open
intervals in $(0, 1)$ with rational endpoints. Since
${\cal B}$ is a countable base, 
we can surely
write
\[
\bigcup_{F \in G({\cal F})}
\bigcup_{i \in {\bf N}} B_i(F) = \bigcup_{j \in {\bf N}}B_j
\]
where all $B_j \in {\cal B}$. 

As shown by Lemma \ref{le-A}, for any $P$-collection $F$, we can
choose $\{I_m\} \in {\cal I}(\Phi(F))$ with
$|I_m| < \epsilon_m$, where $\epsilon_m$ is
sufficiently small for all $m \geq 1$.
Since for any $j \geq 1$, $B_j$ is a subinterval of 
some $I_m \in \{I_m\}$, where
$\{I_m\} \in {\cal I}(\Phi(F))$ for some $P$-collection $F$,
we have $|B_j| \leq |I_m| < \epsilon_m$. We can of course
let $\epsilon_m$ in the above inequality be
less than $2^{-j}\tau$ for any $\tau > 0$.
Consequently,
\[
\bigcup_{F \in G({\cal F})} \Phi(F) \subset \bigcup_{j \in {\bf N}}B_j
\]
and
\[
\sum_{j \in {\bf N}}|B_j| <
\sum_{j \in {\bf N}} 2^{-j}\tau \leq \tau.
\]
The last sum equals $\tau$ if ${\bf N}$ is the set of all
positive integers.
Thus, $\cup_{F \in G({\cal F})} \Phi(F)$ is a set of measure zero.

\subsection*{Version B}
\hskip\parindent
We begin with (\ref{eq-A}) established in Version A.
But we no longer require $B_i(F)$ to be intervals.
For a $P$-collection $F$,
write
\[
C(F) = \bigcup_{i \in {\bf N}} B_i(F)
\]
where $B_i(F)$ does not appear in $C(F^{\prime})$ for
any $P$-collection $F^{\prime} \not = F$.
We list $B_i(F)$ only once in 
$\cup_{F \in G({\cal F})}\cup_{i \in {\bf N}} B_i(F)$.
As a result, any $B_i(F)$ appears only in one $C(F)$.
Since there are at most countably many $B_i(F)$
in $\cup_{F \in G({\cal F})}\cup_{i \in {\bf N}} B_i(F)$.
there are at most countably many $C(F)$. So
we can use $j = 1, 2, \cdots$ to label different $C(F)$,
i.e., for each $C(F)$, there is a unique positive
integer $j$, such that we can denote 
$C(F)$ by $C_j$. Consequently,
\[
\bigcup_{F \in G({\cal F})} C(F) = \bigcup_{j \in N}C_j
\]
and
\[
\bigcup_{F \in G({\cal F})} \Phi(F) \subset \bigcup_{j \in {\bf N}} C_j.
\]

Clearly, each $C_j$ either equals the union
of a sequence of open intervals covering $\Phi(F)$
for some $P$-collection $F$, or equals a subset
of the union of the covering sequence.
Since the total length of the covering intervals can
be less than any $\epsilon > 0$,
the measure of $C_j$, i.e.,
$\mu(C_j)$, can be less than any $\epsilon > 0$.

Moreover, since $C_j$ is a nonempty open
set, we can express $C_j$ as a countable union
of pairwise disjoint open intervals.
Denote by $I_{j, k}$ such intervals. So
\[
C_j = \bigcup_{k \in {\bf N}}I_{j, k}.
\]
Since $\mu(C_j)$ can be less than any $\epsilon > 0$,
$|I_{j, k}|$ can be less than any $\epsilon > 0$
for any $I_{j, k} \subseteq C_j$. Thus, we can
surely let
\[
|I_{j, k}| < 2^{-(j + k)}\tau
\]
for any $\tau > 0$ (see also Lemma \ref{le-A}). Therefore,
\[
\bigcup_{F \in G({\cal F})} \Phi(F) \subset \bigcup_{j \in {\bf N}}
\bigcup_{k \in {\bf N}}I_{j, k}
\]
and
\[
\sum_{j \in {\bf N}}
\sum_{k \in {\bf N}} |I_{j, k}| <
\sum_{j \in {\bf N}}
\sum_{k \in {\bf N}} 2^{-(j + k)}\tau \leq \tau.
\]
The last sum equals $\tau$ if ${\bf N}$ is the set of all
positive integers.
So $\mu(\cup_{F \in G({\cal F})}\Phi(F)) = 0$.

\section{\label{app-2}Hypothetical Nature of ``Continuous Probabilities''}

One might argue that,
with $\Omega, {\cal F}$ and $P$ as
given in Corollary \ref{co-2},
$(\Omega, {\cal F}, P)$
is not only a valid probability space,
but also a counterexample to Theorem \ref{th-1}.
Based on the same argument, one might
even invent various ``counterexamples'' to
Theorem \ref{th-1}. For instance, one might
consider the ``probability space'' of
any ``continuous random variable'' such a
counterexample. 
However,
the reasoning behind
the above argument does not make any sense,
since it is a tautology.
Any ``continuous probability'' is merely an
assumption in the disguise of a definition.

A probability measure $P$ is a function, defined on
${\cal F}$, a $\sigma$-algebra
of subsets of a sample space $\Omega$. 
The set of 
all values of $P$
is the range of $P$.
Clearly, 
the range 
is part of the definition of $P$.

Therefore, if one claims that
the range of
a ``probability measure'' includes
an interval, then this property of 
``continuous probabilities'' is 
actually part of
the definition of the ``probability measure''.
Although one may verify this property against the
definition, the verification is not
a proof of the validity of the
``probability measure'' itself,
since the ``probability measure'' is just
so defined. Any mathematical reasoning, which begins with
a definition and ends merely with the definition,
is nothing but a tautology.

Some arguments in the ``counterexamples'' 
also reflect 
misunderstanding
on measure theory.
The following argument is representative. 
Denote by $\mu$ the Lebesgue measure.
Let $I(a, b)$ represent an
interval on the real line with endpoints
$a$ and $b$, where $a < b$.
Let $r$ be
an arbitrary number in $I(a, b)$, and
denote by $\{r\}$ the set consisting of only
one element $r$. 
The values assigned to $I(a, b)$ and
$\{r\}$ by
the Lebesgue measure are respectively $b - a$
and 0.

The basis of the argument is 
$I(a, b) = \cup_{r \in I(a, b)}\{r\}$.
By letting $a = 0$ and $b = 1$,
one might use
$I(0, 1) = \cup_{r \in I(0, 1)}\{r\}$,
together with
$\mu(I(0, 1)) = 1$ but $m(\{r\}) = 0$ for
any $r \in I(0, 1)$, to construct a ``counterexample'' to
Theorem \ref{th-1}, 
and argue that $\mu(\cup_{r \in I(0, 1)}\{r\}) = 1$
based on $I(0, 1) = \cup_{r \in I(0, 1)}\{r\}$.
With such argument, one might consider
that
$\mu(I(0, 1))$ is an accumulation of $\mu(\{r\})$
for all $r \in I(0, 1)$, i.e.,
$\mu(I(0, 1))$ equals 1 
by means of addition of all $\mu(\{r\})$
rather than by definition.
This is incorrect.

Consider, in general, a measure space
$(I(a, b), {\cal F}, \mu)$, where
${\cal F}$ is a $\sigma$-algebra, and $\mu$ is the
Lebesgue measure. 
In measure theory, countable additivity
\[
\mu(\cup_{i \in N}A_i) = \sum_{i \in N}\mu(A_i)
\]
with pairwise disjoint 
$A_i \in {\cal F}, i = 1, 2, \cdots$ 
cannot be extended to uncountable additivity,
such as 
$\mu(\cup_{r \in I(a, b)}\{r\}) = \sum_{r \in I(a, b)}\mu(\{r\})$.
By definition (e.g., see \cite{Foll})

\[
\sum_{r \in I(a, b)}\mu(\{r\}) =
\sup\left\{\sum_{r \in A}\mu(r):
A \subset I(a, b),\; \mbox{$A$ is finite}\right\} = 0.
\]

The calculated value of 
$\sum_{r \in I(a, b)}\mu(\{r\})$
contradicts the measure value assigned to the interval
$I(a, b)$ according to the definition of Lebesgue measure.
Thus, 
rather than being the result of
summation of
uncountably many zeros, 
it is just so defined 
that
$\mu(I(a, b)) = b - a$. 
Actually, for a measure space, countable union cannot be
extended to uncountable union. In particular,
we have the following result.

\begin{theo}
\label{th-2}
The expression 
$I(a, b) = \cup_{r \in I(a, b)}\{r\}$ is invalid
for Lebesgue measure.
\end{theo}

\proof
Any set $A$ with $\mu(A) > 0$ has a
non-measurable subset.
For example, let $I(a, b) = [0, 1)$.
A non-measurable subset of $[0, 1)$
is given in \cite{Foll}.
If Theorem \ref{th-2} is false, then
$\mu(\cup_{r \in I(a, b)}\{r\}) =
\mu(I(a, b)) = b - a > 0$.
As a result,
there is a non-measurable subset $V$ of
$\cup_{r \in I(a, b)}\{r\}$. 
Define

\[
W = \cup_{r \in I(a, b)}\{r\}\setminus V.
\]

If $W = \emptyset$, 
then 
$I(a, b) = \cup_{r \in I(a, b)}\{r\}$ 
implies that $I(a, b)$ is non-measurable.
This is a contradiction.
So we assume 
$W \not = \emptyset$.
By the well-ordering principle, there is
a (strict) well ordering $\prec$ for $W$ \cite{Foll}. Write

\[
W_r = \{x \in W: x \prec r\}, \; r \in W.
\]
Let $\alpha$ be the first element of $W$. 
Define

\[
E = \{r \in W: \{r\}\cup W_r\;
\mbox{is measurable}\}.
\]
Since $\{\alpha\} \cup W_{\alpha} = 
\{\alpha\}\cup \emptyset = \{\alpha\}$,
and since $\{\alpha\}$ is measurable,
$\alpha \in E$.
There are two cases.

(i) The set $W$ has a last element $\beta$,
and $W_r$ has a last element $\eta(r)$ for 
each $r \in W$.
If $W_r \subset E$, then for any $x \in W_r$,
we have $x \in E$. As a result,
$\{x\}\cup W_x$ is measurable for
any $x \in W_r$. In particular,
$\{\eta(r)\}\cup W_{\eta(r)}$ is
measurable. So

\[
\{r\}\cup W_r = \{r\}\cup(\{\eta(r)\}\cup W_{\eta(r)})
\]
is measurable. Consequently, $r \in E$.
By induction on $W$, we have
$W = E$. Therefore, $\beta \in E$, and hence

\[
\{\beta\}\cup W_{\beta} = W
\]
is measurable. As a result

\[
\cup_{r \in I(a, b)}\{r\}\setminus W = V
\]
is measurable. We see a contradiction again.

(ii) The set $W$, or $W_r$ for some
$r \in W$, does not have a last element.
Define $Z = \{z + d: z \in C\}$,
where $d$ is a constant,
such that $I(a, b)\cap Z = \emptyset$, and
$C$ is the Cantor set. Since $\mu(C) = 0$, and since
Lebesgue measure is translation invariant \cite{Foll}, 
$\mu(Z) = \mu(C) = 0$.

If $W_r$ does not have a last element for
some $r \in W$, then we take an element of $Z$
that has not been
taken for any $W_s, s \not = r$. 
Denote this element of $Z$ also by $\eta(r)$, 
and the set of all such $\eta(r)$ by ${\cal H}$. 
Since ${\cal H} \subset Z$, we have
$\mu({\cal H}) = 0$. 
If $W_r$ has a last element for each $r \in W$, then 
${\cal H} = \emptyset$.

We extend
the order $\prec$ by setting
$\eta(r) \prec r$ and $x \prec \eta(r)$
for all $x \in W_r$.
If $W_s$ and $W_r$ do not have last elements, where
$s\prec r$, 
then with such extension, we have (a)
$\eta(s)\prec\eta(r)$, (b) for any $x \in W_r$,
if $x \prec s$, then $x \prec \eta(s)$;
otherwise $\eta(s) \prec x$, and
(c) $x \prec \eta(r)$ for any $x \in W_s$.
Define 
$U_r = \{\eta(r)\}\cup W_r\cup\{\eta(s) \in {\cal H}: s\prec r\}$. 
For $W_r$ with a
last element, 
$U_r = W_r\cup\{\eta(s) \in {\cal H}: s\prec r\}$.

If $W$ does not have a last element, then
we take an element not in $\cup_{r \in I(a, b)}\{r\}\cup {\cal H}$,
denote this element also by $\beta$, and
define $U = \{\beta\}\cup W\cup {\cal H}$. We further
extend the order $\prec$ by setting
$r \prec \beta$ for all $r \in W\cup {\cal H}$.
If $W$ already has a last element $\beta$, then 
$\beta \in W$, and
$U = W\cup {\cal H}$. It is easy to verify that
$U$ is well ordered by the
extended order $\prec$, and has the same first
element as that of $W$.

Now $U$ has a last element $\beta$, and
$U_r$ has a last element for each $r \in U$,
which is either the last element of $W_r$, or
$\eta(r) \in {\cal H}$.
We use $U$ and $U_r$
to replace $W$ and $W_r$, respectively, and
consider measure space
$(I(a, b)\cup {\cal H} \cup \{\beta\},
{\cal F}\cup {\cal F}({\cal H}, \beta), \mu)$
instead of $(I(a, b), {\cal F}, \mu)$, where
${\cal F}({\cal H}, \beta)$ is the family of all subsets of 
${\cal H}\cup \{\beta\}$.
With the same argument for case (i),

\[
\funbi
{\{\beta\}\cup U_{\beta} = U}
{\{\beta\}\cup W\cup {\cal H},}{$\beta \not\in W$}
{W\cup {\cal H},}{$\beta \in W$}
\]
is measurable. Consequently,

\[
\funbi
{W}{U\setminus (\{\beta\}\cup {\cal H}),}{$\beta \not\in W$}
{U\setminus {\cal H},}{$\beta \in W$}
\]
is measurable. This again leads to the
contradiction in case (i).

\proofend

We can also obtain the above result 
based on a simple observation.
One of the most important notions
in measure theory is that of neglecting
sets of measure zero. After neglecting
sets of measure zero from the
measure space, it can be seen
clearly that
$I(a, b) = \cup_{r \in I(a, b)}\{r\}$ is invalid.

A probability measure is also a measure.
The elucidation above shows clearly why a probability measure must
be defined on a $\sigma$-algebra $\cal F$ for an
uncountable sample space like $I(a, b)$. 
In other words, for such a sample
space $\Omega = I(a, b)$, we must assign probabilities to
subsets of $\Omega$. 

For measure spaces like $(I(a, b), {\cal F}, \mu)$,
a necessary condition is that
${\cal F}$ cannot include all subsets of
$I(a, b)$. This is because some subsets are
not measurable
in the sense of Lebesgue measure.
So the definition of ${\cal F}$ imposes some
restrictions on the members of
${\cal F}$. Yet Theorem \ref{th-1}
shows that such restrictions are not
restrictive enough to make $(I(a, b), {\cal F}, P)$,
where $P = \mu/(b - a)$,
a valid
probability space.
A more stringent restriction that
the range of the probability measure must be
a set of Lebesgue measure zero
should be
imposed on any probability space. 
With this restriction, 
the $\sigma$-algebra ${\cal F}$ of the
probability space
$(I(a, b), {\cal F}, P)$
cannot include
all subintervals of $I(a, b)$.


\section{\label{app-3}
Response to Comment (quant-ph/0509130)
}

This appendix is a point-by-point response to
a comment (quant-ph/0509130) on our paper (quant-ph/0509089).
We find the comment incorrect.

\subsection{\label{sec:level1}
Points Raised in quant-ph/0509130
}

{\em Point 1}: The comment 
(quant-ph/0509130) claims a 
``counterexample'' $(\Omega, {\cal F}, P)$,
where $\Omega = [0, 1], {\cal F} = $ Borel sets in $[0, 1]$,
and $P$ is the Lebesgue measure 
(referred to as Lebesgue-Borel measure in quant-ph/0509130)
restricted to the
$\sigma$-algebra ${\cal F}$ of the Borel sets.
With this ``counterexample'', it is claimed, in quant-ph/0509130,
that $P({\cal F})$ (the range of $P$) is $[0, 1]$.

{\em Point 2}: Let $S = \cup_{F \in G({\cal F})}\Phi(F)$
(the right side is generally defined in our paper quant-ph/0509089).
It is claimed, in quant-ph/0509130, that $S$ is not
necessarily a member of the $\sigma$-algebra ${\cal F}$.

{\em Point 3}: From $P(\Phi(F)) = 0$ for any
$F \in G({\cal F})$, one cannot conclude
$P(S) = \sum_{F \in G({\cal F})}P(\Phi(F)) = 0$.

\subsection{Our Response
}

{\em Response to Point 1}: The ``counterexample'' is 
a meaningless tautology. We have intensively deliberated
on this issue in our paper. Please see Appendix B in
quant-ph/0509089.

Any ``continuous probability'', like $P$ given in the
``counterexample'', is merely an assumption in
the disguise of a definition. The range of $P$ is
part of the definition of $P$. 
Such definition causes contradictions, as
we have proved in
quant-ph/0509089 that the range of a probability measure
cannot include any interval.
The definition of $P$ in the ``counterexample'' also
makes the comment in quant-ph/0509130 self-contradictory.

To see the self-contradiction,
here is an example. If $P$ is a probability
measure, then any element of $S$ is a
value of $P$ in $(0, 1)$. 
Denote by ${\cal R}$
the range of
$P$ after removing 0 and 1. So ${\cal R}$
is a subset of $S$, and $S \subset (0, 1)$.
On the other hand, if the range of $P$ 
is $[0, 1]$ as
claimed in quant-ph/0509130 (Point 1), then
${\cal R} = (0, 1)$ and
$S = (0, 1)$. Clearly, the
open unit interval $(0, 1)$ is 
a member of the
$\sigma$-algebra ${\cal F}$ of the
Borel sets in $[0, 1]$. This contradicts the
claim in Point 2
that $S$ is not necessarily a member of ${\cal F}$.

{\em Response to Point 2}:
We have proved that,
for a probability measure,
$\cup_{F \in G({\cal F})}\Phi(F)$
is a set of Lebesgue measure zero
in quant-ph/0509089. Since Lebesgue measure restricted
to the $\sigma$-algebra of Borel sets is not complete,
a set of Lebesgue measure zero is not necessarily
a member of the $\sigma$-algebra.

{\em Response to Point 3}: Point 3 is misleading,
since our proof of the invalidity of
``continuous probabilities'' 
in quant-ph/0509089
is based on a concept in topology called
countable base, and
does not involve
$P(S) = \sum_{F \in G({\cal F})}P(\Phi(F)) = 0$.
We give a concise version of the proof in
the main text, and two additional versions with
more intensive deliberations in Appendix A.
Point 3 is irrelevant to any
of the three versions of our proof.

\subsection{Conclusion
}

We have responded,
point-by-point, to the comment in quant-ph/0509130
on our paper quant-ph/0509089.
We conclude that the comment is incorrect.


\bibliography{draft-1}

\end{document}